\documentclass[aps,prb,twocolumn,floatfix]{revtex4}

\usepackage{graphicx}
\usepackage{graphics}
\usepackage{amsmath} 

\newcommand{\green}[2]{\ensuremath{\langle\!\langle #1;#2\rangle\!\rangle}}
\newcommand{\corr}[1]{\ensuremath{\langle #1 \rangle}}
\begin{document}

%\preprint{}

\title{The ground state magnetic phase diagram of the ferromagnetic Kondo-lattice model}

\author{S.~Henning}
\email[]{henning@physik.hu-berlin.de}
\author{W.~Nolting}
\affiliation{Lehrstuhl Festk\"orpertheorie, Institut f\"ur Physik, Humboldt-Universit\"at zu Berlin, Newtonstrasse 15, 12489 Berlin, Germany}

\date{\today}

\begin{abstract}
	The magnetic ground state phase diagram of the ferromagnetic Kondo-lattice model is 
	constructed by calculating internal energies of all possible bipartite magnetic
	configurations of the simple cubic lattice explicitly. This is done in one dimension
	(1D), 2D and 3D for a local moment of $S=\frac{3}{2}$. By assuming saturation in the local
	moment system we are able to treat all appearing higher local correlation functions within an 
	equation of motion approach exactly. A simple explanation
	for the obtained phase diagram in terms of bandwidth reduction is given. Regions
	of phase separation are determined from the internal energy curves by an explicit 
	Maxwell construction.
\end{abstract}

%\keywords{}

\maketitle

\section{Introduction}
The ferromagnetic Kondo lattice model (FKLM), also referred to as $s$-$d$ model or 
double exchange model, is the basic model for understanding magnetic phenomena in
systems where local magnetic moments couple ferro-magnetically to itinerant carriers.
This holds for a wide variety of materials. 

In the context of transition metal compounds
Zener proposed the double exchange mechanism to explain ferromagnetic (FM) metallic phase
in the manganites \cite{Zener51_1,Zener51_2}. In these materials the Mn $5d$ shells are split
by the crystal field into three degenerate $t_{2g}$ orbitals which are localized and form
a total spin $S=\frac{3}{2}$ according to atomic selection rules and two $e_g$ orbitals 
providing the itinerant electrons. These electrons couple via Hund exchange coupling 
ferro magnetically with the localized spins. Therefore the FKLM is a basic ingredient 
to describe the rather complex physics of the manganites \cite{Dagotto03, Stier07, Stier08}.

Another nearly ideal field of application of the FKLM is the description of the rare earth materials 
Gd and EuX (X=O,S,Se,Te). These materials have a half filled $4f$ shell in common that is strongly localized
and the electrons in this shell couple to a total spin momentum of $S=\frac{7}{2}$. The FKLM was then used successfully to
explain the famous redshift of the absorption edge of the optical $4f$-$5d$ transition in the
ferromagnetic semiconductor EuO\cite{Busch64,Rys67}. In [\onlinecite{Santos04}] a many-body analysis of the FKLM 
in combination with a band structure calculation was used to get a realistic value for the
Curie temperature of the ferromagnetic metal Gd that is in good agreement with experiment.  

Although it is necessary to extent the FKLM in order to get a realistic description of
the above mentioned examples knowledge of the properties of the pure (single band) FKLM is crucial
for understanding these materials.

To reveal the ground state magnetic phases one has to solve the many-body problem of the FKLM.
This was already done in previous works by using different techniques. Dynamical mean field theory
(DMFT) was used by several authors [\onlinecite{Yunoki98,Kagan99,Chattopadhyay01,Lin05}] to get
information about different magnetic domains. In [\onlinecite{Pekker05}] a continuum field theory
approach was used to derive the 2D phase-diagram at $T=0$. Classical Monte Carlo simulations were
performed in [\onlinecite{Yunoki98, Dagotto98}]. For 1D systems numerical exact density-matrix renormalization
group calculations were done in [\onlinecite{Garcia04}]. In [\onlinecite{Kienert06}] the authors 
have used a Green function method to test the validity of assuming the quantum localized spins
to be classical objects. Extended FKLMs including more material specific effects were for instance 
investigated in [\onlinecite{Peters07,Stier08}].

In this work we will compare all bipartite magnetic configurations for the simple cubic (sc) lattice
by calculating their respective internal energies. To this end the electronic Green function has to
be determined. This is done by an equation of motion approach and, assuming that the local moment
system is saturated, we are able to show that all appearing
local higher correlation functions can be treated exactly. From the calculated internal energies the 
phase-diagram is constructed and region of phase-separation are determined.

The paper is organized as follows. In Sec.~\ref{sec:model&theory} the model Hamiltonian and details
of the calculation are presented. In Sec.~\ref{sec:results&discussion} we discuss the phase-diagrams
and give an explanation for the sequence of phases obtained by looking at the quasi-particle density
of states. In Sec.~\ref{sec:summary&outlook} we summarize the results and give an outlook on possible
directions for further research.

\section{Model and Theory}
\label{sec:model&theory}
\subsection{Model Hamiltonian}
\begin{figure}
	\includegraphics[width=8.0cm]{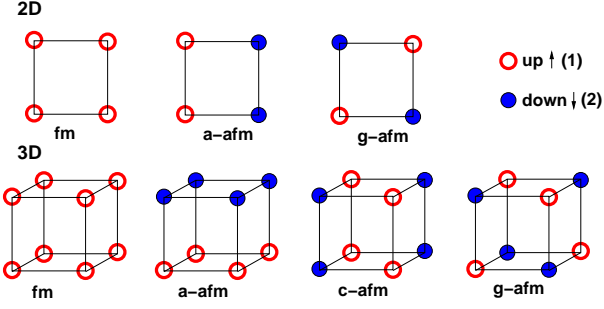}
	\caption{\label{fig:latticetypes}(Color online) Magnetic phases considered in this work (1D omitted).}
\end{figure}

For a proper description of different (anti-) ferromagnetic alignments of localized 
magnetic moments it is useful to divide the full lattice into two or more sub-lattices 
(primitive cells) each ordering ferro magnetically.\\ In this work we only consider simple 
cubic bipartite lattices, i.e. anti-ferromagnetic configurations that can be obtained
by dividing the simple cubic lattice into two sub-lattices.   
In Fig.(\ref{fig:latticetypes}) all possible decompositions in two and three dimensions are shown.
In case of 1D only the ferromagnetic and g-type anti-ferromagnetic phase remain.
The Hamiltonian of the FKLM in second quantization reads as follows:
\begin{eqnarray}
  \label{eq:hamiltonian}
  \lefteqn{H=H_{s}+H_{sf}=\sum_{ij\sigma}\sum_{\alpha \beta}T^{\alpha \beta}_{ij}
  							   c^+_{i\alpha\sigma}c_{j\beta\sigma}}\nonumber\\
    & &   -\frac{J}{2}\sum_{i\sigma}\sum_{\alpha}\left( z_\sigma S^z_{i\alpha}
          c^+_{i\alpha\sigma}c_{i\alpha\sigma}+S^\sigma_{i\alpha}c^+_{i\alpha-\sigma}
          c_{i\alpha\sigma}\right).
\end{eqnarray}
The first term describes the hopping of Bloch electrons with spin $\sigma$ between different 
sites. The lattice sites $\mathbf{R}_{i\alpha}$ are denoted by a Latin index $i$ for 
the unit cell and an Greek index $\alpha \in {A,B}$ for the corresponding sub-lattice, 
i.e. $\mathbf{R}_{i\alpha}=\mathbf{R}_i+\mathbf{r}_{\alpha}$.  
The second term describes a local Heisenberg-like exchange interaction between the itinerant electrons
and local magnetic moments $\mathbf{S}_{i\alpha}$ 
where $J>0$ is the strength of this interaction, $z_{\uparrow\downarrow}=\pm 1$ 
accounts for the two possible spin projections of the electrons and ($S^{\sigma}_{i\alpha}=
S^{x}_{i\alpha} + z_{\sigma}iS^{y}_{i\alpha}$) denotes the spin raising/lowering operator.\\
\subsection{internal energy}
The internal energy of the FKLM at $T=0$ is given by ground state expectation value
of the Hamiltonian:
\begin{equation}
U=\langle H \rangle = \frac{1}{2}\sum_{\alpha\sigma}\int_{-\infty}^{\infty}
	f_{-}(E)ES_{\alpha\sigma}(E)dE
\label{eq:internalE}
\end{equation}
where $S_{\alpha\sigma}(E)=-\frac{1}{\pi}\mathrm{Im}G_{\alpha\sigma}(E)$ is 
the local spectral density, $f_{-}(E)$ denotes the Fermi function and $G_{\alpha\sigma}(E)$ denotes
the local electronic Green function (GF). Note, that this formula is obtained by a straightforward 
calculation of the ground-state expectation value of the Hamiltonian (\ref{eq:hamiltonian})
using the spectral theorem and is therefore exact. 

Our starting point is the equation of motion (EQM) for the electronic GF:
\begin{equation}
	\sum_{l\gamma}\left( E\delta^{\alpha\gamma}_{il}-T^{\alpha\gamma}_{il}\right)
		G^{\gamma\beta}_{lj\sigma} = \delta^{\alpha\beta}_{ij}
	-\frac{J}{2}\left( I^{\alpha\alpha\beta}_{iij\sigma}
	                   + F^{\alpha\alpha\beta}_{iij\sigma}
				\right)
\label{eq:elecGF_EQM}
\end{equation}
with Ising-GF: 
$I^{\alpha\gamma\beta}_{ikj\sigma}=z_{\sigma}\green{S^z_{i\alpha}c_{k\gamma\sigma}}{c^+_{j\beta\sigma}}$
and spin-flip-GF: 
$F^{\alpha\gamma\beta}_{ikj\sigma}=\green{S^{-\sigma}_{i\alpha}c_{k\gamma-\sigma}}{c^+_{j\beta\sigma}}$.
Our basic assumption for the ground state is perfect saturation of the local moment system 
\footnote{Although it is known that the Neel state is not the ground state of a Heisenberg anti-ferromagnet 
deviations from saturation are small for a local magnetic moment $S > \frac{1}{2}$ (see e.g. 
Ref.~[\onlinecite{Anderson52}]).}. With this assumption the Ising-GF can be decoupled exactly:
\begin{equation}
I^{\alpha\gamma\beta}_{ikj\sigma}(E)\rightarrow z_{\sigma}z_{\alpha}SG^{\gamma\beta}_{kj}(E)
\label{eq:ising_decoup}
\end{equation}
where $z_{\alpha}=\pm 1$ denotes the direction of sub-lattice magnetization. 
In a first attempt to solve Eq.~(\ref{eq:elecGF_EQM}) we have neglected spin-flip processes
completely ($F^{\alpha\gamma\beta}_{ikj\sigma}\approx0$). With (\ref{eq:ising_decoup}) we then get
a closed system of equations which can be solved for the electronic GF by Fourier transformation:
\begin{eqnarray}
G^{(\mathrm{MF})}_{\alpha\sigma}(E) 
	&=& \frac{1}{N}\sum_{\mathbf{q}}
		G^{\alpha\alpha(\mathrm{MF})}_{\mathbf{q}\sigma}(E)\\
	&=&	\frac{1}{N}\sum_{\mathbf{q}}\frac{1}
	   	{E+z_{\sigma}z_{\alpha}\frac{J}{2}S-\epsilon^{\alpha\alpha}_{\mathbf{q}}
		-\frac{\epsilon^{\alpha\bar{\alpha}}_{\mathbf{q}}\epsilon^{\bar{\alpha}\alpha}_{\mathbf{q}}}
		      {E+z_{\sigma}z_{\bar{\alpha}}\frac{J}{2}S
					-\epsilon^{\bar{\alpha}\bar{\alpha}}_{\mathbf{q}}}\nonumber
		 }
\label{eq:green_MF}
\end{eqnarray}
where $\epsilon^{\alpha\beta}_{\mathbf{q}}$ is the Fourier transform of the hopping integral
and $\bar{\alpha}=-\alpha$ denotes the complementary sub-lattice. We will call this solution the 
``mean-field'' (MF) solution. Note, that the ferromagnetic phase is contained in the above
formula by setting $\epsilon^{\alpha\bar{\alpha}}_{\mathbf{q}}$ to zero.

To go beyond the MF treatment it is necessary to find a better approximation for
the spin-flip-GF. To this end we write down the EQM for the spin-flip-GF:
\begin{eqnarray}
\label{eq:sf_EQM}
\lefteqn{
	\sum_{l\mu}\left( E\delta^{\gamma\mu}_{kl} - T^{\gamma\mu}_{kl}\right)
		F^{\alpha\mu\beta}_{ilj\sigma} =}\\ 
& 		\green{\left[S^{-\sigma}_{i\alpha},H_{sf}\right]_{-}
	 			c_{k\gamma-\sigma}}{c^+_{j\beta\sigma}}+ 
	 \green{S^{-\sigma}_{i\alpha}\left[c_{k\gamma-\sigma},H_{sf}\right]_{-}}{c^+_{j\beta\sigma}}\nonumber
\end{eqnarray}
Our strategy to get an approximate solution for the spin-flip-GF is to treat the non-local
correlations on a mean-field level whereas the local terms will be treated more carefully.
This is similar to the idea of the dynamical mean field theory (DMFT) developed for strongly
correlated electron systems.\cite{Georges96} Let us start with the non-local 
($i\ne k$ or $i=k$ but $\alpha \ne \gamma$) GFs first. 
It can be shown \cite{Nolting97} that the higher GFs resulting from the commutator 
of $S^{-\sigma}_{i\alpha}$ with $H_{sf}$ are approximately given by the product of the spin-flip-GF 
times spin-wave energies of the local moment system. Therefore it is justified to neglect the 
resulting GFs since the spin-wave energies are typically 3-4 orders of magnitude smaller than 
the local coupling $J$ \cite{Nolting97, Santos02}. \\
The second term on the rhs of (\ref{eq:sf_EQM}) gives two higher GFs which we decouple on a 
mean-field level:
\begin{eqnarray}
\lefteqn{
	\green{S^{-\sigma}_{i\alpha}\left[c_{k\gamma-\sigma},H_{sf}\right]_{-}}{c^+_{j\beta\sigma}} 
	\approx -\frac{J}{2}}\nonumber\\
	&\left(\corr{S^{-\sigma}_{i\alpha}S^{\sigma}_{k\gamma}}
		\green{c_{k\gamma\sigma}}{c^+_{j\beta\sigma}} 
		-z_{\sigma}\corr{S^{z}_{k\gamma}}\green{S^{-\sigma}_{i\alpha}c_{k\gamma-\sigma}}{c^+_{j\beta\sigma}}
			 \right)\nonumber\\
	&\rightarrow z_{\sigma}z_{\gamma}S\frac{J}{2} F^{\alpha\gamma\beta}_{ikj\sigma}.
\label{eq:sf_approx_nonloc}
\end{eqnarray}
where in the last step the saturated sub-lattice magnetization is exploited.\\
We now come to the local terms ($i=k$, $\alpha=\gamma$). The two higher GFs resulting
from the second commutator on the rhs of (\ref{eq:sf_EQM}) reduce to:
\begin{eqnarray}
\green{S^{-\sigma}_{i\alpha}S^{\sigma}_{i\alpha}c_{i\alpha\sigma}}{c^+_{j\beta\sigma}}
&\rightarrow& S(1-z_{\sigma}z_{\alpha})G^{\alpha\beta}_{ij\sigma}\\
\green{S^{-\sigma}_{i\alpha}S^{z}_{i\alpha}c_{i\alpha-\sigma}}{c^+_{j\beta\sigma}}
&\rightarrow& (z_{\alpha}S + z_{\sigma}\delta_{-\sigma\alpha})F^{\alpha\alpha\beta}_{iij\sigma}.\nonumber
\label{eq:loccorr_1}
\end{eqnarray}
Additionally we get a higher order Ising-GF and spin-flip-GF from the first
commutator. The higher order spin-flip-GF can be treated {\it exactly} by using the
EQM of the (known) Ising-GF given in the appendix (\ref{eq:ising_EQM}).
This leads to:
\begin{eqnarray}
\lefteqn{
\green{S^{-\sigma}_{i\alpha}n_{i\alpha\sigma}c_{i\alpha-\sigma}}{c^+_{j\beta\sigma}}
	\rightarrow }\nonumber\\
	&z_{\sigma}z_{\alpha}\frac{2}{J}S\left(\delta^{\alpha\beta}_{ij}-\sum_{l\mu}
			\left( (E+z_{\sigma}z_{\alpha}\frac{J}{2}S)\delta^{\alpha\mu}_{il}-T^{\alpha\mu}_{il}\right)
			G^{\mu\beta}_{lj\sigma}\right)\nonumber\\
	&	-\left(z_{\sigma}z_{\alpha}S-\delta_{\sigma\alpha}\right)F^{\alpha\alpha\beta}_{iij\sigma}.
\label{eq:higher_sf}
\end{eqnarray}
The higher order Ising-GF can be traced back to the higher order spin-flip-GF by
writing down its EQM and make use of saturation in the local-moment system (see appendix
\ref{app:higherIsing} for details):
\begin{eqnarray}
\lefteqn{
\green{S^{z}_{i\alpha}n_{i\alpha-\sigma}c_{i\alpha\sigma}}{c^+_{j\beta\sigma}}
\rightarrow z_{\alpha}S\left(G^{\alpha\beta(\mathrm{MF})}_{ij\sigma}
		\corr{n_{j\beta-\sigma}}\right.}\nonumber\\
	&\left. -\frac{J}{2}\sum_{l\gamma}G^{\alpha\gamma(\mathrm{MF})}_{il\sigma}
		\green{S^{-\sigma}_{l\gamma}n_{l\gamma\sigma}c_{l\gamma-\sigma}}{c^+_{j\beta\sigma}}\right).
\label{eq:higher_ising}
\end{eqnarray}
It is a major result of this work that it is possible to incorporate all
local correlations without approximation, i.e. to treat all local 
higher order GFs exactly.
Combining the results for the appearing higher GFs found in (\ref{eq:sf_approx_nonloc}), 
(\ref{eq:loccorr_1}), (\ref{eq:higher_sf}) and (\ref{eq:higher_ising}) we can
now solve (\ref{eq:sf_EQM}) for the spin-flip-GF: 
\begin{widetext}
\begin{equation}
F^{\alpha\alpha\beta}_{iij\sigma}=-\frac{JSG^{(\mathrm{MF})}_{\alpha-\sigma}}
										{1+z_{\sigma}z_{\alpha}\frac{J}{2}G^{(\mathrm{MF})}_{\alpha-\sigma}}
	\left(z_{\sigma}z_{\alpha}G^{\alpha\beta(\mathrm{MF})}_{ij\sigma}\left(\corr{n^{\beta}_{j-\sigma}}-\delta_{\sigma\beta}\right)
		  +\sum_{l\gamma}\left(\delta^{\alpha\gamma}_{il}\delta_{\sigma-\alpha}+
		  	G^{\alpha\gamma(\mathrm{MF})}_{il\sigma}\delta_{\sigma\gamma}\sum_{t\eta}
			\left( G^{\mu\nu(\mathrm{MF})}_{jk\sigma} \right)^{-1\:\gamma\eta}_{lt}\right)
		  	G^{\eta\beta}_{tj\sigma}
	\right).
\end{equation}
Inserting this result into (\ref{eq:elecGF_EQM}) and performing a Fourier transformation 
we finally get:
\begin{equation}
\label{eq:electGF_sf}
	\sum_{\gamma}\left(\left(G^{\mu\nu(\mathrm{MF})}_{\mathbf{q}\sigma}\right)^{-1}_{\alpha\gamma} 
		-A^\alpha_\sigma\left(\delta_{\sigma-\alpha}\delta_{\alpha\gamma}+
			G^{\alpha\sigma(\mathrm{MF})}_{\mathbf{q}\sigma}
			\left(G^{\mu\nu(\mathrm{MF})}_{\mathbf{q}\sigma}\right)^{-1}_{\sigma\gamma}\right)\right)
			G^{\gamma\beta}_{\mathbf{q}\sigma}(E) = \delta_{\alpha\beta} +
			z_{\sigma}z_{\alpha}A^\alpha_\sigma G^{\alpha\beta(\mathrm{MF})}_{\mathbf{q}\sigma}
			\left(\corr{n^{\beta}_{-\sigma}}-\delta_{\sigma\beta}\right)
\end{equation}
\end{widetext}
with
\begin{equation*}
A^\alpha_\sigma(E) = \frac{J^2 S G^{(\mathrm{MF})}_{\alpha-\sigma}(E)}
		  		          {2+z_{\sigma}z_{\alpha}JG^{(\mathrm{MF})}_{\alpha-\sigma}(E)}.
\end{equation*}
This equation allows for a self-consistent calculation of the electronic GF
and we will call this the spin-flip (SF) solution.\\
One important test for the above result is to compare it with exact known
limiting cases. We found that (\ref{eq:electGF_sf}) reproduces the solution of
the ferro-magnetically saturated semiconductor \cite{Shastry81,Allan82} in the 
limit of zero band-occupation. Additionally the 4-peak structure of the 
spectrum as known from the ``zero-bandwidth''-limit \cite{Nolting84} is retained
whereas the peaks are broadened to bands with their center of gravity at the original
peak positions.
\subsection{phase separation}
\begin{figure}
	\includegraphics[width=6.0cm,height=4.0cm]{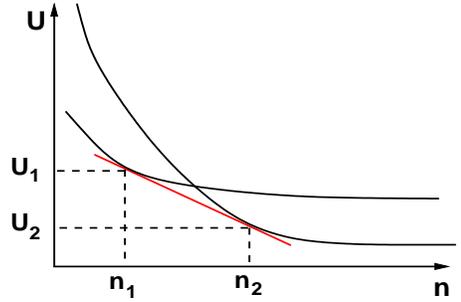}
	\caption{\label{fig:maxwell}(Color online) Explicit Maxwell construction for determining the
	boundaries of phase separated regions.}
\end{figure}

To determine the regions of phase separation in the phase diagram we have
used an explicit Maxwell construction as shown in Fig.\ref{fig:maxwell}.
The condition for the boundaries of the phase separated region is:
\begin{equation}
	\left.\frac{dU_1}{dn}\right|_{n=n_1}=\frac{U_2(n_2)-U_1(n_1)}{n_2-n_1}=
	\left.\frac{dU_2}{dn}\right|_{n=n_2}.
	\label{eq:maxwell}
\end{equation}

\section{Results and Discussion}
\label{sec:results&discussion}
\begin{figure}
	\includegraphics[width=8.0cm]{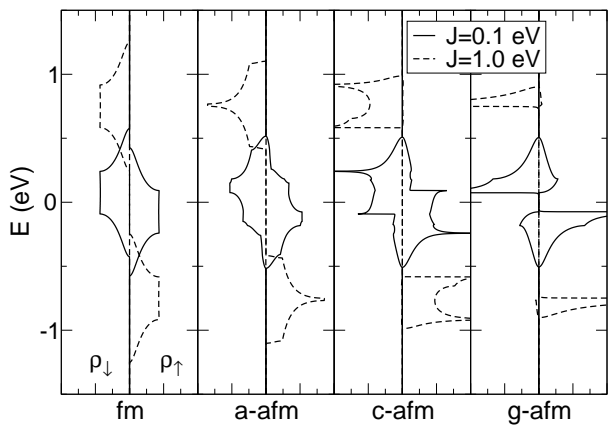}
	\caption{\label{fig:dos_MF}Sub-lattice quasi particle density of states (QDOS) of up and down
	electrons obtained from the MF-GF (\ref{eq:green_MF}) for two values of local coupling $J$ 
	shown for different magnetic configurations. 
	Parameters: $S=\frac{3}{2}$ and free electron bandwidth: $W=1.0$ eV.}
\end{figure}

The internal energy of the FKLM at $T=0$ is given as an integral (\ref{eq:internalE}) over the product
of (sub-lattice) quasi-particle density of states (QDOS) times energy up to Fermi-energy.
For understanding the resulting phase-diagrams it is therefore useful to have a closer look at the
QDOS first. In Fig.\ref{fig:dos_MF} the sub-lattice MF-QDOS is shown for the different magnetic phases investigated
(in 3D).
The underlying full lattice is of simple cubic type with nearest neighbor hopping $T$ chosen such that
the bandwidth $W$ is equal to $W=1$ eV in the case of free electrons ($J=0$ eV). 
The local magnetic moment is equal to $S=\frac{3}{2}$.
We have plotted the up and down-electron spectrum separately for two different values of 
$J=0.1/1.0$ eV.
The exchange splitting $\Delta_{ex}=JS$ eV of up and down-band is clearly visible.
The decisive difference between the phases for nonzero values of $J$ is bandwidth reduction
from ferromagnetic over a, c to g-afm phase. The reason for this behavior becomes clear by
looking at the magnetic lattices shown in Fig.\ref{fig:latticetypes}. In the 
ferromagnetic case an (up-)electron can move freely in all 3 directions of space without
paying any additional potential energy. In a-type anti-ferromagnetic phase the electron 
can still move freely within a plane but when moving in the direction perpendicular to the plane
it needs to overcome an energy-barrier $\Delta_{ex}$. Hence the QDOS for large values of $J$
resembles the form of 2D tight-binding dispersion. The bandwidth is reduced due to the 
confinement of the electrons. In the c-afm phase the electron can only move freely along one 
direction and the QDOS becomes effectively one dimensional. Finally in the g-type phase the
electron in the large $J$ limit is quasi localized and the bandwidth gets very small.
We will see soon that this bandwidth-effect is mainly responsible for the structure of
the phase-diagram. Before we come to this point we want to discuss the influence of
spin-flip processes as incorporated in (\ref{eq:electGF_sf}).
\begin{figure}
	\includegraphics[width=8.0cm]{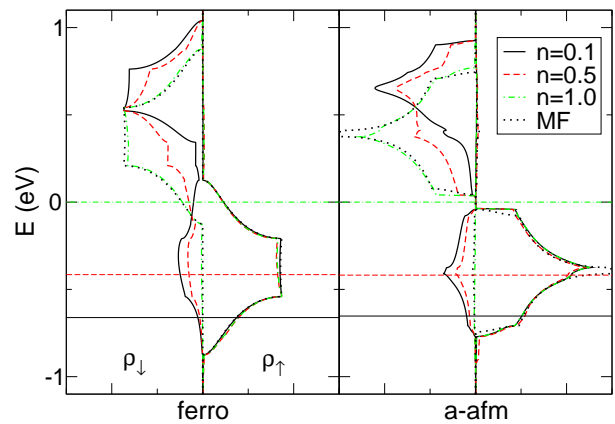}
	\caption{\label{fig:dos_MCDA}(Color online) Sub-lattice QDOS of up and down electrons obtained from the
	SF-GF (\ref{eq:electGF_sf}) for three different band-fillings $n$ shown for the
	ferromagnetic and a-afm phase. The local coupling $J=0.5$ eV is fixed. Dotted line:
	corresponding MF result. Horizontal lines: respective Fermi levels. Other parameters as in Fig.\ref{fig:dos_MF}.}
\end{figure}
In Fig.\ref{fig:dos_MCDA} the QDOS for $J=0.5$ eV is shown for three different 
band fillings $n$. The corresponding Fermi energies are marked by horizontal lines. 
The apparent new feature are the scattering states in the down spectrum for
band fillings below half filling. Thereby the spectral weight of the scattering states 
is more and more reduced with increasing Fermi level. 
A second effect is that the sharp features in the 
MF-QDOS of the anti-ferromagnetic phases are smeared out. Compared to the MF results
the overall change of QDOS below Fermi energy due to the inclusion of spin-flip 
processes is small and will not affect the
form of the phase-diagram drastically. However non-negligible changes can be expected.
Note that the model shows perfect particle-hole symmetry. Therefore the results for the
internal energy will be the same for $n=x$ and $n=2-x$ ($x=0\dots1$, $n=1$: half filling).\\
We come now to the discussion of the phase-diagrams which we got by comparing the internal energies
of the different phases explicitly.
\begin{figure*}
	\includegraphics[width=16.0cm]{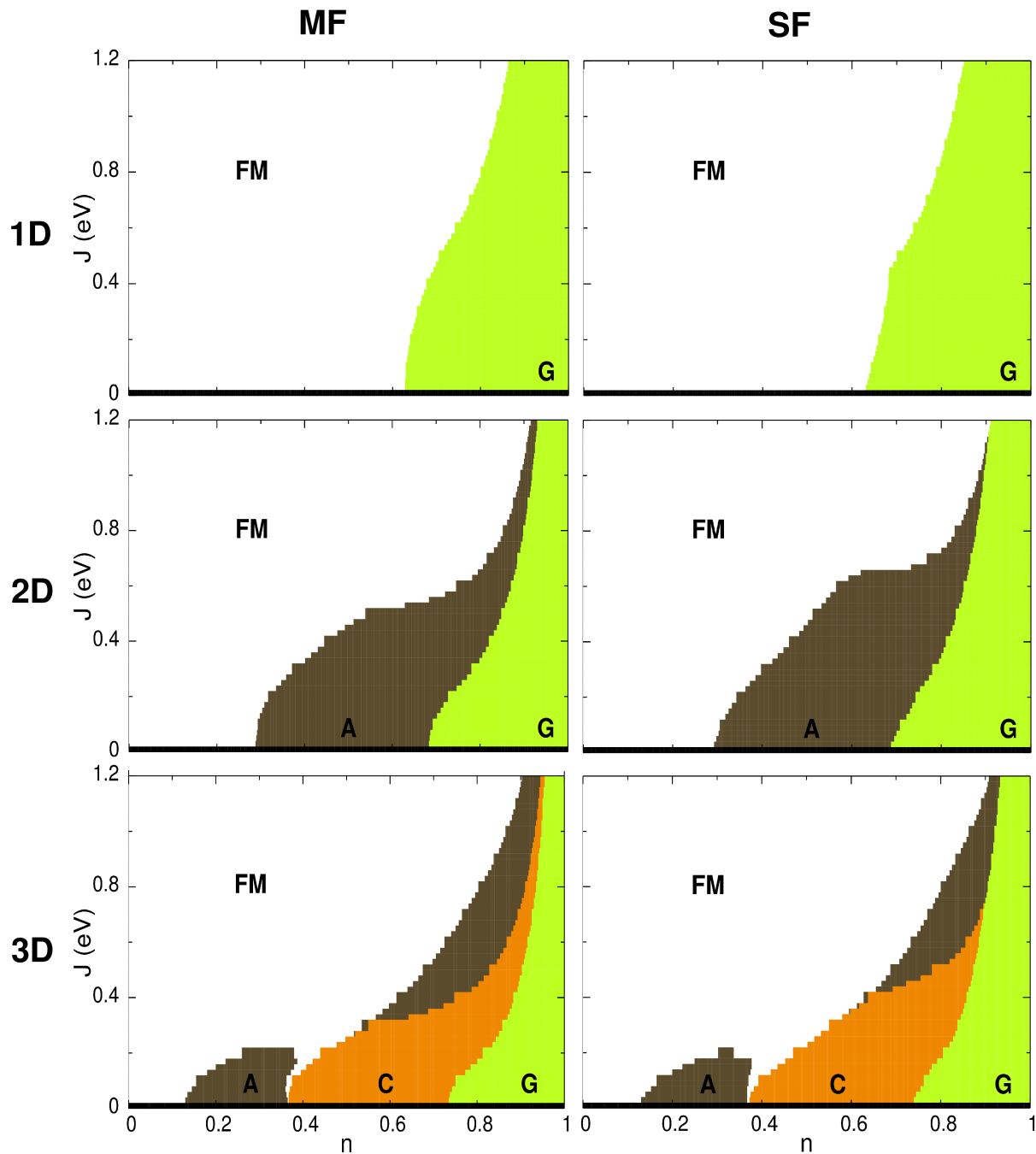}
	\caption{\label{fig:phase_os}(Color online) First column: Phase-diagram as function
	of band-filling $n$ and local coupling $J$ obtained with
	mean-field (MF) theory (\ref{eq:green_MF}) in one, two and three dimensions. Second column:
	Phase-diagram obtained by inclusion of spin-flip processes (SF) (\ref{eq:electGF_sf}).
	Regions of different colors mark different (magnetic) phases: ferromagnetic (white),
	a-afm (brown), c-afm (orange), g-afm (green) and paramagnetic phase (black).}
\end{figure*}
The pure phase-diagrams (without phase-separation) are shown in Fig.\ref{fig:phase_os} whereas the 
different phases are marked by color code. In the first column
the results of the MF-calculation are shown for the 1-, 2- and 3-dimensional case. The second
column shows the effects of inclusion of spin-flip processes. We will concentrate here mainly
onto the 3D case since most of the given arguments hold equally for the 1D and 2D case. 
For $J=0$ the system is paramagnetic (black bar at bottom). For larger $J$ ($J>0$) a typical sequence 
appear: for low band-fillings $n$ the system is always ferromagnetic and, with increasing $n$,
it becomes a-type then c-type and finally g-type anti-ferromagnetic. This behavior is 
understood easily by looking at the formula for the internal energy (\ref{eq:internalE}) and the
MF-QDOS in Fig.\ref{fig:dos_MF}. Because of the bandwidth-effect discussed already the
band-edge of the ferromagnetic state is always lowest in energy and will give therefore the
lowest internal energy for small band-occupation. But since the QDOS of the anti-ferromagnetic
phases increase much more rapidly than the ferromagnetic one, these give             
more weight to low energies in the integral (\ref{eq:internalE}) and will become lowest in energy
eventually for larger band-fillings. Therefore the bandwidth-effect is main effect explaining
the order of phases with increasing $n$. A very interesting feature can be found
in the region: $J=0.2\dots0.3$. In this region the ferromagnetic phase is directly followed by
the c-afm phase for increasing $n$ although the a-afm phase has a larger bandwidth than the c-afm phase.
This can be explained by the two-peak structure of the c-afm-QDOS. Due to the first peak at low energies 
these energies are much more weighted than in the a-afm case and the c-afm phase will become lower
in energy than the a-afm phase. Since the reduction of bandwidth of the anti-ferromagnetic phases
compared to the ferromagnetic phase is more pronounced for larger values of $J$ the ferromagnetic
region is growing in this direction.\\ The paramagnetic phase (black bar at $J=0$) disappear 
for any finite $J$ since due to the down-shift of the up-spectrum of the ferromagnetic phase 
their internal energy will always be lower.
\begin{figure*}
	\includegraphics[width=16.0cm]{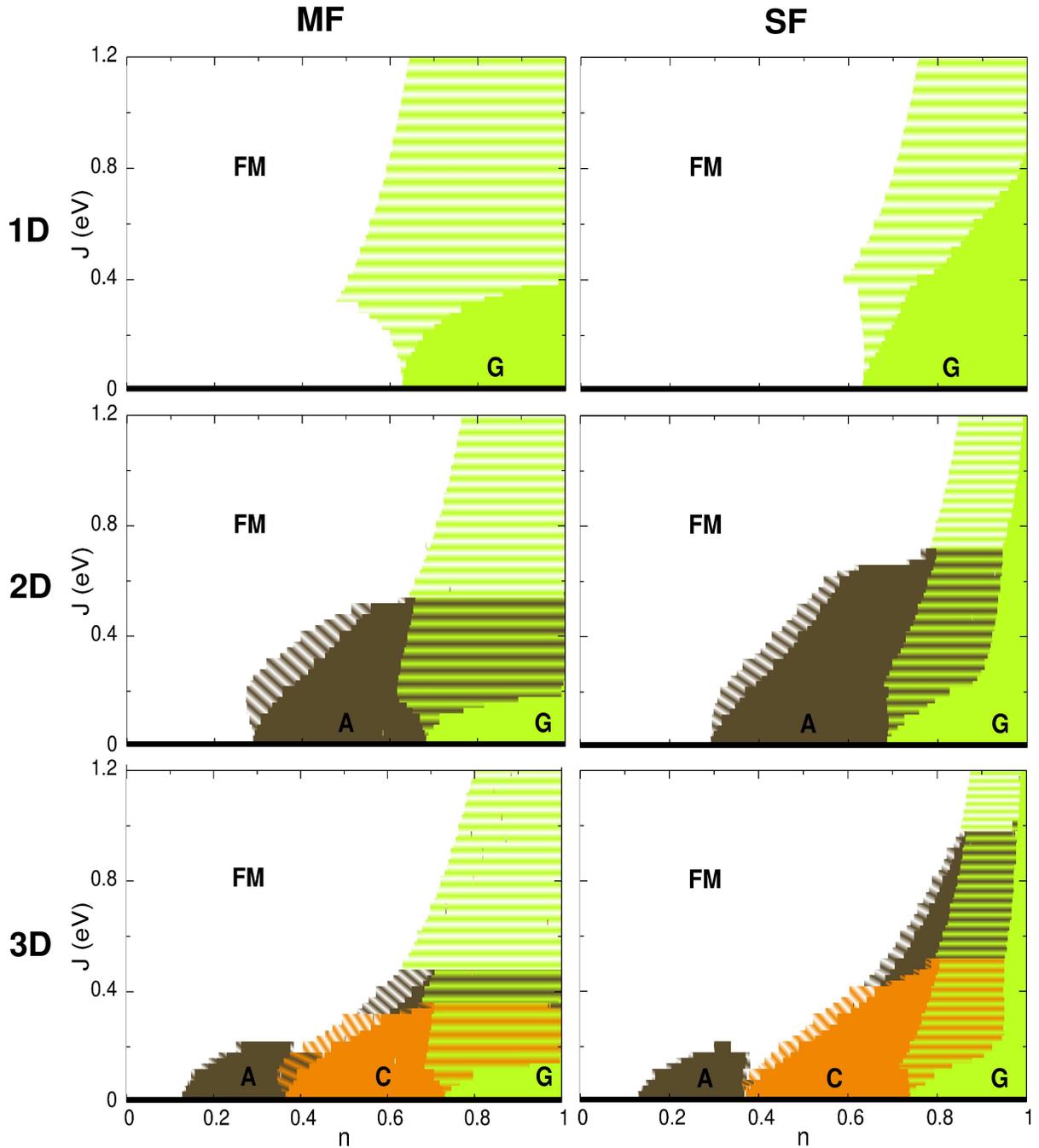}
	\caption{\label{fig:phase_ms}(Color online) Phase-diagram with phase-separation. Regions
	of phase-separation are marked with two-colored stripes. Color code as in Fig.\ref{fig:phase_os}.}
\end{figure*}
When comparing the MF and the SF-phase-diagram they appear to be
very similar at first glance. However two interesting differences can be found, namely an 
increased $J$ region without a-afm-phase and the vanishing c-phase above $J \approx 0.8$eV.

Fig.\ref{fig:phase_ms} shows the phase-diagrams where regions of phase-separation, which we have
determined by an explicit Maxwell construction (\ref{eq:maxwell}), are marked by colored stripes.
The two colors denote the involved pure phases. As one can see large regions become phase-separated,
whereas the two participating phases are mostly determined by the adjacent pure phases. There is one
interesting exception from this: above a certain $J$ only fm/g-afm phase-separation survives and
suppresses all other phases in this area. Inclusion of spin-flip processes as shown in the right
column of Fig.\ref{fig:phase_ms} push this $J$ up to higher values. Generally spin-flip
processes seem to reduce phase-separation as can be seen in the g-afm phase and e.g. at the
border between fm and c-afm phase.

Our results are in good qualitative agreement with numerical and DMFT results reported by 
others\cite{Dagotto98,Chattopadhyay01,Lin05}. It is common to all these works that for small 
coupling strength $J$ there is only a small ferromagnetic region at low band occupation $n$
followed by more complicated (anti-ferromagnetic, spiral, canted) spin states/phase-separation.
With increasing $J$ the region of fm is also increased to larger $n$ values. Near half-filling
($n=1$) one will find always anti-ferromagnetism/phase-separation.   phase-diagram very similar to our
2D-FM result shown in Fig.\ref{fig:phase_ms} was obtained by Pekker et.al.[\onlinecite{Pekker05}].
The positions of A and G phase are in nearly perfect agreement. However the authors seem not to have 
taken into account phase-separation between A and G phase and their finding of FM/A phase-separation near half-filling at larger $J$ is not in accordance with our results.
\section{Summary and Outlook}
\label{sec:summary&outlook}
We have constructed phase diagrams of the FKLM in 1D, 2D and 3D by comparing the internal energies of all
possible bipartite magnetic configurations of the simple cubic lattice. To this end the 
electronic GF is calculated by an EQM approach. We can show, that it is possible to treat all
appearing higher local correlation functions exact and we derive an explicit formula for
the electronic GF (\ref{eq:electGF_sf}). The obtained sequence of phases with increasing band
occupation $n$ and Hunds coupling $J$ is explained by the
reduction of QDOS bandwidth due to electron confinement. Region of phase separation are then determined
from the internal energy curves by an explicit Maxwell construction.

In the phase diagram obtained only phases appear that have explicitly considered by us. Therefore an 
important extension of this work could be the inclusion of more complicated spin structures like
canted/spiral spin states as reported by others [\onlinecite{Pekker05,Garcia04}]. However the
bandwidth criterion obtained here can certainly be applied to such more complicated states also.

\appendix
\section{EQM of the Ising-GF}
\begin{eqnarray}
\lefteqn{\sum_{l\mu}\left( E\delta^{\gamma\mu}_{kl} - T^{\gamma\mu}_{kl}\right)
			I^{\alpha\mu\beta}_{ilj\sigma} = 
			z_{\sigma}\delta^{\gamma\beta}_{kj}\langle S^z_{i\alpha} \rangle }\nonumber\\
 & & 
  	 -\frac{J}{2}\left(\green{S^z_{i\alpha}S^z_{k\gamma}c_{k\gamma\sigma}}{c^+_{j\beta\sigma}}
 	 +z_{\sigma}
	 \green{S^z_{i\alpha}S^{-\sigma}_{k\gamma}c_{k\gamma-\sigma}}{c^+_{j\beta\sigma}}\right. \nonumber\\
 & & \left. +z_{\sigma}\sum_{\sigma'}z_{\sigma'}
 	 \green{S^{\sigma'}_{i\alpha}c^+_{i\alpha-\sigma'}c_{i\alpha\sigma'}c_{k\gamma\sigma}}
	 	   {c^+_{j\beta\sigma}}\right),
\label{eq:ising_EQM}
\end{eqnarray}

\section{higher order Ising-GF}
\label{app:higherIsing}
The higher order Ising-GF can be decomposed into:
\begin{equation}
\green{S^{z}_{i\sigma}n_{i\alpha-\sigma}c_{i\alpha\sigma}}{c^{+}_{j\beta\sigma}}\rightarrow
z_{\alpha}S\green{n_{i\alpha-\sigma}c_{i\alpha\sigma}}{c^{+}_{j\beta\sigma}}
\end{equation} 
when a saturated sub-lattice magnetization is assumed. The EQM of the remaining GF turns out to be:
\begin{eqnarray}
\label{eq:EQM_higherIsing}
\lefteqn{(E+z_{\sigma}z_{\alpha}\frac{J}{2}S)\green{n_{i\alpha-\sigma}c_{i\alpha\sigma}}
												   {c^{+}_{j\beta\sigma}}} \nonumber\\
	=& \sum_{l\gamma}T^{\alpha\gamma}_{il}\green{c^+_{l\gamma-\sigma}c_{i\alpha-\sigma}c_{i\alpha\sigma}}
												   {c^{+}_{j\beta\sigma}}  &(\mathrm{I})\nonumber\\
	+& \sum_{l\gamma}T^{\alpha\gamma}_{il}\green{c^+_{i\alpha-\sigma}c_{l\gamma-\sigma}c_{i\alpha\sigma}}
												   {c^{+}_{j\beta\sigma}}  &(\mathrm{II})\nonumber\\
	+& \sum_{l\gamma}T^{\alpha\gamma}_{il}\green{c^+_{i\alpha-\sigma}c_{i\alpha-\sigma}c_{l\gamma\sigma}}
												   {c^{+}_{j\beta\sigma}}  &(\mathrm{III})\nonumber\\
	-& 2\sum_{l\gamma}T^{\alpha\gamma}_{il}\green{c^+_{l\gamma-\sigma}c_{i\alpha-\sigma}c_{i\alpha\sigma}}
												   {c^{+}_{j\beta\sigma}}  &\nonumber\\
	+& \delta^{\alpha\beta}_{ij}\corr{n_{i\alpha-\sigma}} 
		-\frac{J}{2}\green{S^{-\sigma}_{i\alpha}n_{i\alpha\sigma}c_{i\alpha-\sigma}}{c^{+}_{j\beta\sigma}}.&
\end{eqnarray}
Subtracting the term denoted by (I) from this equation one gets: 
\begin{eqnarray}
  \lefteqn{\sum_{l\gamma}(E\delta^{\alpha\gamma}_{il}-T^{\alpha\gamma}_{il}+z_{\sigma}z_{\alpha}\delta^{\alpha\gamma}_{il}
			\frac{J}{2}S)\times}\nonumber\\
	& &	\green{c^+_{l\gamma-\sigma}c_{i\alpha-\sigma}c_{i\alpha\sigma}}{c^{+}_{j\beta\sigma}}\nonumber\\
	&=& \sum_{l\gamma}T^{\alpha\gamma}_{il}\green{c^+_{i\alpha-\sigma}c_{l\gamma-\sigma}c_{i\alpha\sigma}}
												   {c^{+}_{j\beta\sigma}}  \nonumber\\
	&+& \sum_{l\gamma}T^{\alpha\gamma}_{il}\green{c^+_{i\alpha-\sigma}c_{i\alpha-\sigma}c_{l\gamma\sigma}}
												   {c^{+}_{j\beta\sigma}}  \nonumber\\
	&-& 2\sum_{l\gamma}T^{\alpha\gamma}_{il}\green{c^+_{l\gamma-\sigma}c_{i\alpha-\sigma}c_{i\alpha\sigma}}
												   {c^{+}_{j\beta\sigma}}  \nonumber\\
	&+& \delta^{\alpha\beta}_{ij}\corr{n_{i\alpha-\sigma}} 
		-\frac{J}{2}\green{S^{-\sigma}_{i\alpha}n_{i\alpha\sigma}c_{i\alpha-\sigma}}
												   {c^{+}_{j\beta\sigma}}.
\end{eqnarray}
This can be solved for $\green{n_{i\alpha-\sigma}c_{i\alpha\sigma}}{c^{+}_{j\beta\sigma}}$ 
by left-multiplying with the MF-GF matrix: 
\begin{eqnarray}
\label{eq:1_sol}
\lefteqn{\green{n_{i\alpha-\sigma}c_{i\alpha\sigma}}{c^{+}_{i\alpha\sigma}}=} \nonumber\\
	& & \sum_{kl\eta\gamma}G^{(\mathrm{MF})\alpha\eta}_{ik\sigma}T^{\eta\gamma}_{kl}
		\green{c^+_{i\alpha-\sigma}c_{l\gamma-\sigma}c_{i\alpha\sigma}}{c^{+}_{j\beta\sigma}}  \nonumber\\
	&+& \sum_{kl\eta\gamma}G^{(\mathrm{MF})\alpha\eta}_{ik\sigma}T^{\eta\gamma}_{kl}
		\green{c^+_{i\alpha-\sigma}c_{i\alpha-\sigma}c_{l\gamma\sigma}}{c^{+}_{j\beta\sigma}}  \nonumber\\
	&-& 2\sum_{kl\eta\gamma}G^{(\mathrm{MF})\alpha\eta}_{ik\sigma}T^{\eta\gamma}_{kl}
		\green{c^+_{l\gamma-\sigma}c_{i\alpha-\sigma}c_{i\alpha\sigma}}{c^{+}_{j\beta\sigma}}  \nonumber\\
	&+& G^{(\mathrm{MF})\alpha\beta}_{ij\sigma}\corr{n_{j\beta-\sigma}}		\nonumber \\ 
	&-& \frac{J}{2}\sum_{k\eta}G^{(\mathrm{MF})\alpha\eta}_{ik\sigma}
		\green{S^{-\sigma}_{k\eta}n_{k\eta\sigma}c_{k\eta-\sigma}}{c^{+}_{j\beta\sigma}}.
\end{eqnarray}
Two other equations are obtained from (\ref{eq:EQM_higherIsing}) by subtracting term (II) or (III)
and performing the same steps as before. This yields:
\begin{eqnarray}
\label{eq:2_sol}
\lefteqn{\green{n_{i\alpha-\sigma}c_{i\alpha\sigma}}{c^{+}_{i\alpha\sigma}}=} \nonumber\\
	& & \sum_{kl\eta\gamma}G^{(\mathrm{MF})\alpha\eta}_{ik\sigma}T^{\eta\gamma}_{kl}
		\green{c^+_{l\gamma-\sigma}c_{i\alpha-\sigma}c_{i\alpha\sigma}}{c^{+}_{j\beta\sigma}}  \nonumber\\
	&+& \sum_{kl\eta\gamma}G^{(\mathrm{MF})\alpha\eta}_{ik\sigma}T^{\eta\gamma}_{kl}
		\green{c^+_{i\alpha-\sigma}c_{i\alpha-\sigma}c_{l\gamma\sigma}}{c^{+}_{j\beta\sigma}}  \nonumber\\
	&-& 2\sum_{kl\eta\gamma}G^{(\mathrm{MF})\alpha\eta}_{ik\sigma}T^{\eta\gamma}_{kl}
		\green{c^+_{l\gamma-\sigma}c_{i\alpha-\sigma}c_{i\alpha\sigma}}{c^{+}_{j\beta\sigma}}  \nonumber\\
	&+& G^{(\mathrm{MF})\alpha\beta}_{ij\sigma}\corr{n_{j\beta-\sigma}} 	\nonumber\\
	&-& \frac{J}{2}\sum_{k\eta}G^{(\mathrm{MF})\alpha\eta}_{ik\sigma}
		\green{S^{-\sigma}_{k\eta}n_{k\eta\sigma}c_{k\eta-\sigma}}{c^{+}_{j\beta\sigma}}
\end{eqnarray}
and
\begin{eqnarray}
\label{eq:3_sol}
\lefteqn{\green{n_{i\alpha-\sigma}c_{i\alpha\sigma}}{c^{+}_{i\alpha\sigma}}=} \nonumber\\
	& & \sum_{kl\eta\gamma}G^{(\mathrm{MF})\alpha\eta}_{ik\sigma}T^{\eta\gamma}_{kl}
		\green{c^+_{l\gamma-\sigma}c_{i\alpha-\sigma}c_{i\alpha\sigma}}{c^{+}_{j\beta\sigma}}  \nonumber\\
	&+& \sum_{kl\eta\gamma}G^{(\mathrm{MF})\alpha\eta}_{ik\sigma}T^{\eta\gamma}_{kl}
		\green{c^+_{i\alpha-\sigma}c_{l\gamma-\sigma}c_{i\alpha\sigma}}{c^{+}_{j\beta\sigma}}  \nonumber\\
	&-& 2\sum_{kl\eta\gamma}G^{(\mathrm{MF})\alpha\eta}_{ik\sigma}T^{\eta\gamma}_{kl}
		\green{c^+_{l\gamma-\sigma}c_{i\alpha-\sigma}c_{i\alpha\sigma}}{c^{+}_{j\beta\sigma}}  \nonumber\\
	&+& G^{(\mathrm{MF})\alpha\beta}_{ij\sigma}\corr{n_{j\beta-\sigma}} 	\nonumber\\
	&-& \frac{J}{2}\sum_{k\eta}G^{(\mathrm{MF})\alpha\eta}_{ik\sigma}
		\green{S^{-\sigma}_{k\eta}n_{k\eta\sigma}c_{k\eta-\sigma}}{c^{+}_{j\beta\sigma}}
\end{eqnarray}
Adding (\ref{eq:2_sol}) and (\ref{eq:3_sol}) and subtracting (\ref{eq:1_sol}) one finally gets:
\begin{eqnarray}
\lefteqn{
\green{S^{z}_{i\alpha}n_{i\alpha-\sigma}c_{i\alpha\sigma}}{c^+_{j\beta\sigma}}
= z_{\alpha}S\left(G^{\alpha\beta(\mathrm{MF})}_{ij\sigma}
		\corr{n_{j\beta-\sigma}}\right.}\nonumber\\
	&\left. -\frac{J}{2}\sum_{l\gamma}G^{\alpha\gamma(\mathrm{MF})}_{il\sigma}
		\green{S^{-\sigma}_{l\gamma}n_{l\gamma\sigma}c_{l\gamma-\sigma}}{c^+_{j\beta\sigma}}\right).
\end{eqnarray}

%\bibliography{citations}

\begin{thebibliography}{24}
\expandafter\ifx\csname natexlab\endcsname\relax\def\natexlab#1{#1}\fi
\expandafter\ifx\csname bibnamefont\endcsname\relax
  \def\bibnamefont#1{#1}\fi
\expandafter\ifx\csname bibfnamefont\endcsname\relax
  \def\bibfnamefont#1{#1}\fi
\expandafter\ifx\csname citenamefont\endcsname\relax
  \def\citenamefont#1{#1}\fi
\expandafter\ifx\csname url\endcsname\relax
  \def\url#1{\texttt{#1}}\fi
\expandafter\ifx\csname urlprefix\endcsname\relax\def\urlprefix{URL }\fi
\providecommand{\bibinfo}[2]{#2}
\providecommand{\eprint}[2][]{\url{#2}}

\bibitem[{\citenamefont{Zener}(1951{\natexlab{a}})}]{Zener51_1}
\bibinfo{author}{\bibfnamefont{C.}~\bibnamefont{Zener}},
  \bibinfo{journal}{Phys. Rev.} \textbf{\bibinfo{volume}{81}},
  \bibinfo{pages}{440} (\bibinfo{year}{1951}{\natexlab{a}}).

\bibitem[{\citenamefont{Zener}(1951{\natexlab{b}})}]{Zener51_2}
\bibinfo{author}{\bibfnamefont{C.}~\bibnamefont{Zener}},
  \bibinfo{journal}{Phys. Rev.} \textbf{\bibinfo{volume}{82}},
  \bibinfo{pages}{403} (\bibinfo{year}{1951}{\natexlab{b}}).

\bibitem[{\citenamefont{Dagotto}(2003)}]{Dagotto03}
\bibinfo{author}{\bibfnamefont{E.}~\bibnamefont{Dagotto}},
  \emph{\bibinfo{title}{Nanoscale Phase Separation and Colossal
  Magnetoresistance}} (\bibinfo{publisher}{Springer,~Berlin},
  \bibinfo{year}{2003}).

\bibitem[{\citenamefont{{Stier} and {Nolting}}(2007)}]{Stier07}
\bibinfo{author}{\bibfnamefont{M.}~\bibnamefont{{Stier}}} \bibnamefont{and}
  \bibinfo{author}{\bibfnamefont{W.}~\bibnamefont{{Nolting}}},
  \bibinfo{journal}{\prb} \textbf{\bibinfo{volume}{75}},
  \bibinfo{pages}{144409} (\bibinfo{year}{2007}).

\bibitem[{\citenamefont{{Stier} and {Nolting}}(2008)}]{Stier08}
\bibinfo{author}{\bibfnamefont{M.}~\bibnamefont{{Stier}}} \bibnamefont{and}
  \bibinfo{author}{\bibfnamefont{W.}~\bibnamefont{{Nolting}}},
  \bibinfo{journal}{\prb} \textbf{\bibinfo{volume}{78}},
  \bibinfo{pages}{144425} (\bibinfo{year}{2008}).

\bibitem[{\citenamefont{{Busch} et~al.}(1964)\citenamefont{{Busch}, {Junod},
  and {Wachter}}}]{Busch64}
\bibinfo{author}{\bibfnamefont{G.}~\bibnamefont{{Busch}}},
  \bibinfo{author}{\bibfnamefont{P.}~\bibnamefont{{Junod}}}, \bibnamefont{and}
  \bibinfo{author}{\bibfnamefont{P.}~\bibnamefont{{Wachter}}},
  \bibinfo{journal}{Physics Letters} \textbf{\bibinfo{volume}{12}},
  \bibinfo{pages}{11} (\bibinfo{year}{1964}).

\bibitem[{\citenamefont{{Rys} et~al.}(1967)\citenamefont{{Rys}, {Helman}, and
  {Baltensperger}}}]{Rys67}
\bibinfo{author}{\bibfnamefont{F.}~\bibnamefont{{Rys}}},
  \bibinfo{author}{\bibfnamefont{J.~S.} \bibnamefont{{Helman}}},
  \bibnamefont{and}
  \bibinfo{author}{\bibfnamefont{W.}~\bibnamefont{{Baltensperger}}},
  \bibinfo{journal}{Physik der Kondensierten Materie, Volume 6, Issue 2,
  pp.105-125} \textbf{\bibinfo{volume}{6}}, \bibinfo{pages}{105}
  (\bibinfo{year}{1967}).

\bibitem[{\citenamefont{Santos et~al.}(2004)\citenamefont{Santos, Nolting, and
  Eyert}}]{Santos04}
\bibinfo{author}{\bibfnamefont{C.}~\bibnamefont{Santos}},
  \bibinfo{author}{\bibfnamefont{W.}~\bibnamefont{Nolting}}, \bibnamefont{and}
  \bibinfo{author}{\bibfnamefont{V.}~\bibnamefont{Eyert}},
  \bibinfo{journal}{Phys. Rev. B} \textbf{\bibinfo{volume}{69}},
  \bibinfo{pages}{214412} (\bibinfo{year}{2004}).

\bibitem[{\citenamefont{Yunoki et~al.}(1998)\citenamefont{Yunoki, Hu, Malvezzi,
  Moreo, Furukawa, and Dagotto}}]{Yunoki98}
\bibinfo{author}{\bibfnamefont{S.}~\bibnamefont{Yunoki}},
  \bibinfo{author}{\bibfnamefont{J.}~\bibnamefont{Hu}},
  \bibinfo{author}{\bibfnamefont{A.~L.} \bibnamefont{Malvezzi}},
  \bibinfo{author}{\bibfnamefont{A.}~\bibnamefont{Moreo}},
  \bibinfo{author}{\bibfnamefont{N.}~\bibnamefont{Furukawa}}, \bibnamefont{and}
  \bibinfo{author}{\bibfnamefont{E.}~\bibnamefont{Dagotto}},
  \bibinfo{journal}{Phys. Rev. Lett.} \textbf{\bibinfo{volume}{80}},
  \bibinfo{pages}{845} (\bibinfo{year}{1998}).

\bibitem[{\citenamefont{{Kagan} et~al.}(1999)\citenamefont{{Kagan}, {Khomskii},
  and {Mostovoy}}}]{Kagan99}
\bibinfo{author}{\bibfnamefont{M.~Y.} \bibnamefont{{Kagan}}},
  \bibinfo{author}{\bibfnamefont{D.~I.} \bibnamefont{{Khomskii}}},
  \bibnamefont{and} \bibinfo{author}{\bibfnamefont{M.~V.}
  \bibnamefont{{Mostovoy}}}, \bibinfo{journal}{European Physical Journal B}
  \textbf{\bibinfo{volume}{12}}, \bibinfo{pages}{217} (\bibinfo{year}{1999}),
  \eprint{arXiv:cond-mat/9804213}.

\bibitem[{\citenamefont{{Chattopadhyay}
  et~al.}(2001)\citenamefont{{Chattopadhyay}, {Millis}, and {Das
  Sarma}}}]{Chattopadhyay01}
\bibinfo{author}{\bibfnamefont{A.}~\bibnamefont{{Chattopadhyay}}},
  \bibinfo{author}{\bibfnamefont{A.~J.} \bibnamefont{{Millis}}},
  \bibnamefont{and} \bibinfo{author}{\bibfnamefont{S.}~\bibnamefont{{Das
  Sarma}}}, \bibinfo{journal}{\prb} \textbf{\bibinfo{volume}{64}},
  \bibinfo{pages}{012416} (\bibinfo{year}{2001}),
  \eprint{arXiv:cond-mat/0004151}.

\bibitem[{\citenamefont{{Lin} and {Millis}}(2005)}]{Lin05}
\bibinfo{author}{\bibfnamefont{C.}~\bibnamefont{{Lin}}} \bibnamefont{and}
  \bibinfo{author}{\bibfnamefont{A.~J.} \bibnamefont{{Millis}}},
  \bibinfo{journal}{\prb} \textbf{\bibinfo{volume}{72}},
  \bibinfo{pages}{245112} (\bibinfo{year}{2005}),
  \eprint{arXiv:cond-mat/0509004}.

\bibitem[{\citenamefont{{Pekker} et~al.}(2005)\citenamefont{{Pekker},
  {Mukhopadhyay}, {Trivedi}, and {Goldbart}}}]{Pekker05}
\bibinfo{author}{\bibfnamefont{D.}~\bibnamefont{{Pekker}}},
  \bibinfo{author}{\bibfnamefont{S.}~\bibnamefont{{Mukhopadhyay}}},
  \bibinfo{author}{\bibfnamefont{N.}~\bibnamefont{{Trivedi}}},
  \bibnamefont{and} \bibinfo{author}{\bibfnamefont{P.~M.}
  \bibnamefont{{Goldbart}}}, \bibinfo{journal}{\prb}
  \textbf{\bibinfo{volume}{72}}, \bibinfo{pages}{075118}
  (\bibinfo{year}{2005}).

\bibitem[{\citenamefont{Dagotto et~al.}(1998)\citenamefont{Dagotto, Yunoki,
  Malvezzi, Moreo, Hu, Capponi, Poilblanc, and Furukawa}}]{Dagotto98}
\bibinfo{author}{\bibfnamefont{E.}~\bibnamefont{Dagotto}},
  \bibinfo{author}{\bibfnamefont{S.}~\bibnamefont{Yunoki}},
  \bibinfo{author}{\bibfnamefont{A.~L.} \bibnamefont{Malvezzi}},
  \bibinfo{author}{\bibfnamefont{A.}~\bibnamefont{Moreo}},
  \bibinfo{author}{\bibfnamefont{J.}~\bibnamefont{Hu}},
  \bibinfo{author}{\bibfnamefont{S.}~\bibnamefont{Capponi}},
  \bibinfo{author}{\bibfnamefont{D.}~\bibnamefont{Poilblanc}},
  \bibnamefont{and} \bibinfo{author}{\bibfnamefont{N.}~\bibnamefont{Furukawa}},
  \bibinfo{journal}{Phys. Rev. B} \textbf{\bibinfo{volume}{58}},
  \bibinfo{pages}{6414} (\bibinfo{year}{1998}).

\bibitem[{\citenamefont{Garcia et~al.}(2004)\citenamefont{Garcia, Hallberg,
  Alascio, and Avignon}}]{Garcia04}
\bibinfo{author}{\bibfnamefont{D.~J.} \bibnamefont{Garcia}},
  \bibinfo{author}{\bibfnamefont{K.}~\bibnamefont{Hallberg}},
  \bibinfo{author}{\bibfnamefont{B.}~\bibnamefont{Alascio}}, \bibnamefont{and}
  \bibinfo{author}{\bibfnamefont{M.}~\bibnamefont{Avignon}},
  \bibinfo{journal}{Phys. Rev. Lett.} \textbf{\bibinfo{volume}{93}},
  \bibinfo{pages}{177204} (\bibinfo{year}{2004}).

\bibitem[{\citenamefont{{Kienert} and {Nolting}}(2006)}]{Kienert06}
\bibinfo{author}{\bibfnamefont{J.}~\bibnamefont{{Kienert}}} \bibnamefont{and}
  \bibinfo{author}{\bibfnamefont{W.}~\bibnamefont{{Nolting}}},
  \bibinfo{journal}{\prb} \textbf{\bibinfo{volume}{73}},
  \bibinfo{pages}{224405} (\bibinfo{year}{2006}),
  \eprint{arXiv:cond-mat/0606485}.

\bibitem[{\citenamefont{{Peters} and {Pruschke}}(2007)}]{Peters07}
\bibinfo{author}{\bibfnamefont{R.}~\bibnamefont{{Peters}}} \bibnamefont{and}
  \bibinfo{author}{\bibfnamefont{T.}~\bibnamefont{{Pruschke}}},
  \bibinfo{journal}{\prb} \textbf{\bibinfo{volume}{76}},
  \bibinfo{pages}{245101} (\bibinfo{year}{2007}), \eprint{0707.0277}.

\bibitem[{\citenamefont{Georges et~al.}(1996)\citenamefont{Georges, Kotliar,
  Krauth, and Rozenberg}}]{Georges96}
\bibinfo{author}{\bibfnamefont{A.}~\bibnamefont{Georges}},
  \bibinfo{author}{\bibfnamefont{G.}~\bibnamefont{Kotliar}},
  \bibinfo{author}{\bibfnamefont{W.}~\bibnamefont{Krauth}}, \bibnamefont{and}
  \bibinfo{author}{\bibfnamefont{M.~J.} \bibnamefont{Rozenberg}},
  \bibinfo{journal}{Rev. Mod. Phys.} \textbf{\bibinfo{volume}{68}},
  \bibinfo{pages}{13} (\bibinfo{year}{1996}).

\bibitem[{\citenamefont{Nolting et~al.}(1997)\citenamefont{Nolting, Rex, and
  {Mathi~Jaya}}}]{Nolting97}
\bibinfo{author}{\bibfnamefont{W.}~\bibnamefont{Nolting}},
  \bibinfo{author}{\bibfnamefont{S.}~\bibnamefont{Rex}}, \bibnamefont{and}
  \bibinfo{author}{\bibfnamefont{S.}~\bibnamefont{{Mathi~Jaya}}},
  \bibinfo{journal}{J.~Phys.:~Condens.~Matter} \textbf{\bibinfo{volume}{9}},
  \bibinfo{pages}{1301} (\bibinfo{year}{1997}).

\bibitem[{\citenamefont{Santos and Nolting}(2002)}]{Santos02}
\bibinfo{author}{\bibfnamefont{C.}~\bibnamefont{Santos}} \bibnamefont{and}
  \bibinfo{author}{\bibfnamefont{W.}~\bibnamefont{Nolting}},
  \bibinfo{journal}{Phys. Rev. B} \textbf{\bibinfo{volume}{65}},
  \bibinfo{pages}{144419} (\bibinfo{year}{2002}).

\bibitem[{\citenamefont{Shastry and Mattis}(1981)}]{Shastry81}
\bibinfo{author}{\bibfnamefont{B.~S.} \bibnamefont{Shastry}} \bibnamefont{and}
  \bibinfo{author}{\bibfnamefont{D.~C.} \bibnamefont{Mattis}},
  \bibinfo{journal}{Phys. Rev. B} \textbf{\bibinfo{volume}{24}},
  \bibinfo{pages}{5340} (\bibinfo{year}{1981}).

\bibitem[{\citenamefont{{Allan} and {Edwards}}(1982)}]{Allan82}
\bibinfo{author}{\bibfnamefont{S.~R.} \bibnamefont{{Allan}}} \bibnamefont{and}
  \bibinfo{author}{\bibfnamefont{D.~M.} \bibnamefont{{Edwards}}},
  \bibinfo{journal}{J. Phys. C} \textbf{\bibinfo{volume}{15}},
  \bibinfo{pages}{2151} (\bibinfo{year}{1982}).

\bibitem[{\citenamefont{Nolting and Matlak}(1984)}]{Nolting84}
\bibinfo{author}{\bibfnamefont{W.}~\bibnamefont{Nolting}} \bibnamefont{and}
  \bibinfo{author}{\bibfnamefont{M.}~\bibnamefont{Matlak}},
  \bibinfo{journal}{Phys.~Status~Solidi~B} \textbf{\bibinfo{volume}{123}},
  \bibinfo{pages}{155} (\bibinfo{year}{1984}).

\bibitem[{\citenamefont{Anderson}(1952)}]{Anderson52}
\bibinfo{author}{\bibfnamefont{P.~W.} \bibnamefont{Anderson}},
  \bibinfo{journal}{Phys.~Rev.} \textbf{\bibinfo{volume}{86}},
  \bibinfo{pages}{694} (\bibinfo{year}{1952}).

\end{thebibliography}

\end{document}